\documentclass[manuscript,screen,nonacm]{acmart}
\setcopyright{none}
\title{Polylab: A MATLAB Toolbox for Multivariate Polynomial Modeling}
\author{Yi-Shuai Niu}
\affiliation{%
  \institution{Beijing Institute of Mathematical Sciences and Applications (BIMSA)}
  \city{Beijing}
  \country{China}}
\email{niuyishuai@bimsa.cn}
\author{Shing-Tung Yau}
\affiliation{%
  \institution{Yau Mathematical Sciences Center, Tsinghua University}
  \city{Beijing}
  \country{China}}
\affiliation{%
  \institution{Beijing Institute of Mathematical Sciences and Applications (BIMSA)}
  \city{Beijing}
  \country{China}}
\email{styau@tsinghua.edu.cn}

\begin{document}

\begin{abstract}
Polylab is a MATLAB toolbox for multivariate polynomial scalars and polynomial matrices with a unified symbolic-numeric interface across CPU and GPU-oriented backends. The software exposes three aligned classes: \texttt{MPOLY} for CPU execution, \texttt{MPOLY\_GPU} as a legacy GPU baseline, and \texttt{MPOLY\_HP} as an improved GPU-oriented implementation. Across these backends, Polylab supports polynomial construction, algebraic manipulation, simplification, matrix operations, differentiation, Jacobian and Hessian construction, LaTeX export, CPU-side LaTeX reconstruction, backend conversion, and interoperability with YALMIP and SOSTOOLS. Versions~3.0 and~3.1 add two practically important extensions: explicit variable identity and naming for safe mixed-variable expression handling, and affine-normal direction computation via automatic differentiation, MF-logDet-Exact, and MF-logDet-Stochastic. The toolbox has already been used successfully in prior research applications, and Polylab Version~3.1 adds a new geometry-oriented computational layer on top of a mature polynomial modeling core. This article documents the architecture and user-facing interface of the software, organizes its functionality by workflow, presents representative MATLAB sessions with actual outputs, and reports reproducible benchmarks. The results show that MPOLY is the right default for lightweight interactive workloads, whereas MPOLY-HP becomes advantageous for reduction-heavy simplification and medium-to-large affine-normal computation; the stochastic log-determinant variant becomes attractive in larger sparse regimes under approximation-oriented parameter choices.
\end{abstract}


\maketitle

\section{Introduction}
Multivariate polynomials provide a compact modeling language for optimization, control, algebraic geometry, and computational mathematics. Yet software support for them remains awkward because polynomial objects sit at the boundary between symbolic and numerical computation. Users need data structures for polynomial scalars and matrices, overloaded operators that behave naturally in MATLAB, differentiation and evaluation tools, and backend choices that let the same model move from CPU execution to GPU-oriented execution. They also need a reliable way to combine expressions created at different times and in different variable spaces without silently corrupting the model through positional mismatches in exponent storage.

Polylab addresses this software problem directly. It is a general-purpose polynomial toolbox built around three aligned classes: \texttt{MPOLY} (CPU), \texttt{MPOLY\_GPU} (legacy GPU), and \texttt{MPOLY\_HP} (high-performance GPU-oriented). The package supports polynomial construction, algebraic manipulation, simplification, substitution, differentiation, Jacobian and Hessian construction, LaTeX export and CPU-side reconstruction, backend conversion, and bridges to YALMIP and SOSTOOLS. Rather than serving only as a display or parsing layer, it aims to provide an end-to-end symbolic-numeric environment for polynomial workflows inside MATLAB.

Two recent releases substantially extend the scope of the toolbox. Version~3.0 introduces explicit variable identity and naming through \texttt{vars.id} and \texttt{vars.name}. This resolves a persistent fragility of positional polynomial representations, in which expressions created independently can be combined incorrectly if their variable orderings are not aligned. Version~3.1 adds precision control and affine-normal direction computation. The latter connects to the affine-invariant optimization viewpoint developed in Yau's Affine Normal Descent (YAND)~\cite{niu2026yand} and follows the log-determinant geometry framework developed by Niu, Sheshmani, and Yau~\cite{niu2026affinenormal}. Together, these additions move Polylab beyond basic manipulation toward variable-safe modeling and geometry-aware computation.

Polylab complements rather than replaces mathematical software such as YALMIP~\cite{lofberg2004yalmip}, GloptiPoly~3~\cite{henrion2009gloptipoly}, and SOSTOOLS~\cite{papachristodoulou2013sostools}. These packages focus on optimization modeling, moment relaxations, or sum-of-squares workflows. MATLAB's Symbolic Math Toolbox occupies a different position, emphasizing exact symbolic manipulation. Polylab instead focuses on the software infrastructure of polynomial objects themselves: representing them, transforming them, differentiating them, exchanging them, and accelerating them across backends.

The package also arrives with an application record. Earlier versions of Polylab have already been used in Boolean polynomial optimization~\cite{niu2022boolean}, difference-of-SOS and difference-of-convex-SOS decomposition~\cite{niu2024dsos}, and high-order moment portfolio optimization~\cite{niu2019portfolio}. These studies show that the toolbox has already been exercised in combinatorial optimization, algebraic decomposition, and quantitative finance settings where large structured polynomials are central.

This paper makes four contributions. First, it documents the architecture and public functionality of Polylab as a coherent software interface rather than as an isolated collection of methods. Second, it explains the variable-identity mechanism introduced in Version~3.0 and its effect on modeling reliability. Third, it shows how affine-normal computation is exposed in Version~3.1 and how the implemented methods relate to the recent log-determinant framework. Fourth, it reports reproducible benchmarks that separate CPU-favorable and GPU-favorable regimes, making backend selection a concrete user guideline rather than an informal rule of thumb.

The remainder of the article is organized as follows. Section~2 summarizes the package architecture and feature set. Section~3 presents the main polynomial workflows, including variable-safe algebra, differentiation, precision control, and LaTeX round-tripping. Section~4 discusses interoperability with YALMIP, SOSTOOLS, and MATLAB's Symbolic Math Toolbox. Section~5 introduces affine-normal computation and its user-facing interface. Section~6 reports reproducible experimental results, and Sections~7--8 close with discussion, limitations, and conclusions.

\section{Package Overview}
\subsection{Main Classes}
Polylab is organized around three main MATLAB classes.
\begin{itemize}
\item \texttt{MPOLY}: the CPU reference implementation and the semantic baseline of the toolbox.
\item \texttt{MPOLY\_GPU}: a legacy GPU class that mirrors the core interface and serves as a stable GPU baseline. In figures and tables we denote it by MPOLY-GPU.
\item \texttt{MPOLY\_HP}: an improved GPU-oriented class that adds optimized paths for simplification, multiplication, reduction-style operations, and affine-normal computation. In figures and tables we denote it by MPOLY-HP.
\end{itemize}

At the user level, the three classes deliberately look similar. Construction, overloaded algebra, differentiation, and backend conversion are designed to use the same conceptual interface, which reduces the cost of moving between prototyping and performance-oriented execution.

\subsection{Public Functionality by Category}
Table~\ref{tab:featurecatalog} summarizes the public functionality exposed by the current code base and exercised by the bundled tests and tutorials, organized as an interface-level view rather than as a flat list of methods.

\begin{table}
\caption{Public functionality of Polylab grouped by usage scenario.}
\label{tab:featurecatalog}
\begin{tabular}{p{0.23\linewidth}p{0.70\linewidth}}
\toprule
Category & Representative functionality \\
\midrule
Construction and bases & constructor, \texttt{mpolyvars}, \texttt{monolist}, \texttt{zeros}, \texttt{ones}, \texttt{identity}, \texttt{diag}, \texttt{trace}, \texttt{iszero} \\
Algebra and reductions & \texttt{plus}, \texttt{minus}, \texttt{mtimes}, \texttt{times}, \texttt{mpower}, \texttt{power}, \texttt{sum}, \texttt{prod}, \texttt{mrdivide}, \texttt{rdivide}, \texttt{simplify} \\
Inspection and transforms & \texttt{degree}, \texttt{coefficients}, \texttt{mono}, \texttt{eval}, \texttt{subs}, \texttt{quadprod}, \texttt{transmatconvex} \\
Differentiation & \texttt{diff}, \texttt{jacobian}, Hessians via repeated Jacobians \\
Formatting and exchange & \texttt{disp}, \texttt{sdisp}, \texttt{latex}, \texttt{fromlatex} (CPU), \texttt{version}, global precision control \\
Backend conversion & \texttt{MPOLY\_GPU.gpulize}, \texttt{MPOLY\_GPU.cpulize}, \texttt{MPOLY\_HP.gpulize}, \texttt{MPOLY\_HP.cpulize} \\
Affine-normal geometry & \texttt{affineNormalDirection} with \texttt{ad}, \texttt{MF-logDet-Exact}, and \texttt{MF-logDet-Stochastic} methods \\
External interfaces & \texttt{mpoly2yalmip}, \texttt{yalmip2mpoly}, \texttt{yalmip2mpolygpu}, \texttt{yalmip2mpolyhp}, \texttt{mpoly2sostools} \\
\bottomrule
\end{tabular}
\Description{A feature table organizing the public Polylab interface into construction, algebra, inspection, differentiation, formatting, backend conversion, affine-normal geometry, and external interfaces.}
\end{table}

This feature catalog was assembled from the actual class files, the package tutorial, and the test scripts \texttt{test\_MPOLY.m}, \texttt{test\_MPOLY\_GPU.m}, \texttt{test\_MPOLY\_HP.m}, \texttt{tutorial\_MPOLY.m}, and the benchmark scripts added in Version~3.1.

\subsection{Application Footprint}
Table~\ref{tab:applications} summarizes three representative studies that used earlier versions of Polylab in concrete polynomial workflows.

\begin{table}
\caption{Representative application studies that used Polylab.}
\label{tab:applications}
\begin{tabular}{p{0.22\linewidth}p{0.28\linewidth}p{0.42\linewidth}}
\toprule
Area & Representative study & Role of Polylab in the workflow \\
\midrule
Boolean optimization & Niu and Glowinski~\cite{niu2022boolean} & construction and manipulation of Boolean polynomial objectives and update maps in a discrete dynamical systems approach \\
Polynomial decomposition & Niu, Le Thi, and Pham~\cite{niu2024dsos} & large-scale polynomial modeling, algebraic transforms, and differentiation inside difference-of-SOS / difference-of-convex-SOS decomposition workflows \\
Quantitative finance & Niu et al.~\cite{niu2019portfolio} & higher-order moment polynomial modeling and exchange with sums-of-squares and optimization layers for portfolio problems \\
\bottomrule
\end{tabular}
\Description{A table listing representative application studies that used Polylab in Boolean optimization, polynomial decomposition, and high-order portfolio optimization.}
\end{table}

These papers demonstrate application value beyond synthetic microbenchmarks. The affine-normal module introduced in Version~3.1 is therefore added on top of a toolbox that has already seen use in distinct large-scale polynomial settings.

The next sections illustrate the main calling patterns. Table~\ref{tab:apisignature} gives a compact signature-level guide to the most important public entry points.

\begin{table}
\caption{Representative API signatures, inputs, and outputs used throughout the toolbox.}
\label{tab:apisignature}
\begin{tabular}{p{0.28\linewidth}p{0.28\linewidth}p{0.33\linewidth}}
\toprule
Representative call & Typical input & Output / effect \\
\midrule
\texttt{x = MPOLY.mpolyvars(n,'x')} & nonnegative integer \texttt{n}, optional base name & \texttt{n x 1} variable vector with \texttt{vars.id} and \texttt{vars.name} \\
\texttt{p = MPOLY(n,coef,pow)} & coefficient vector and exponent matrix with matching monomial count & scalar polynomial with \texttt{n} variables \\
\texttt{q = simplify(p,zeroprec)} & polynomial scalar or matrix, optional zero threshold & same-size polynomial with merged monomials and small terms removed \\
\texttt{g = jacobian(p)} and \texttt{H = jacobian(g)} & scalar polynomial & gradient row and Hessian-like polynomial array \\
\texttt{v = p.eval(x0)} and \texttt{ps = p.subs(expr)} & numeric point or replacement polynomial vector & numeric value or substituted polynomial object \\
\texttt{MPOLY.setEvalBackend(mode)} and \texttt{MPOLY.getEvalBackend()} & \texttt{'auto'}, \texttt{'mex'}, or \texttt{'matlab'} & selects the default CPU evaluation path for double-precision inputs \\
\texttt{s = latex(p,'x','\%.6e')} and \texttt{p2 = MPOLY.fromlatex(s)} & polynomial scalar/matrix or LaTeX string & formatted string or reconstructed CPU polynomial \\
\texttt{pg = MPOLY\_HP.gpulize(p)} and \texttt{pc = MPOLY\_HP.cpulize(pg)} & CPU or HP polynomial object & converted object on target backend \\
\texttt{[d,info] = affineNormalDirection(...)} & scalar polynomial, evaluation point, option struct & affine-normal direction plus diagnostics \\
\bottomrule
\end{tabular}
\Description{A table summarizing representative Polylab API signatures for variable creation, construction, simplification, differentiation, evaluation, LaTeX exchange, backend conversion, and affine-normal computation.}
\end{table}

Throughout the paper we use short MATLAB-console transcripts prefixed by \texttt{>>}. They show the concrete calling rules and selected outputs needed to understand the interface. Full function-by-function documentation remains in the repository docstrings and tutorial scripts.

\begin{table}
\caption{Representative operator families and helper semantics.}
\label{tab:coresemantics}
\begin{tabular}{p{0.33\linewidth}p{0.58\linewidth}}
\toprule
Function family & Meaning in Polylab \\
\midrule
\texttt{monolist(n,d)} & returns the column vector of all monomials in $n$ variables of total degree at most $d$; useful as a basis generator in approximation, SOS, and moment-style workflows \\
\texttt{zeros}, \texttt{ones}, \texttt{identity} & create constant zero, one, and identity polynomial arrays with a prescribed variable space \\
\texttt{diag}, \texttt{trace}, \texttt{iszero} & build or extract diagonals, sum diagonal entries, and test whether a polynomial scalar or matrix is structurally zero \\
\texttt{plus}, \texttt{minus}, \texttt{times}, \texttt{rdivide}, \texttt{power} & MATLAB-style elementwise arithmetic on polynomial objects, with variable-space alignment performed automatically when needed \\
\texttt{mtimes}, \texttt{mrdivide}, \texttt{mpower}, \texttt{sum}, \texttt{prod} & matrix multiplication, right division, matrix power, and dimensionwise reductions for polynomial arrays \\
\texttt{simplify} & merges repeated monomials and removes coefficients below a user-chosen tolerance \\
\texttt{degree}, \texttt{coefficients}, \texttt{mono} & inspect total degree, retrieve coefficient lists, and extract selected monomials with unit leading coefficients \\
\texttt{eval}, \texttt{subs} & evaluate at a numeric point or substitute polynomial expressions for variables \\
\texttt{quadprod}, \texttt{transmatconvex} & provide optimization-oriented constructions: fast quadratic forms $b^\top Q b$ and basis transforms for homogeneous convex polynomial representations \\
\bottomrule
\end{tabular}
\Description{A table summarizing the semantics of major operator families and helper functions in Polylab, including basis generation, matrix helpers, arithmetic, simplification, structural inspection, evaluation, substitution, and optimization-oriented transforms.}
\end{table}

This semantic summary highlights the role of each major function family, while exhaustive argument-by-argument detail remains in the bundled tutorial and docstrings.

\section{Core Polynomial Workflows}
\subsection{Construction, Display, and Basic Queries}
The fundamental user entry point is \texttt{mpolyvars}, which returns a column vector of scalar variables with attached metadata. A scalar polynomial can then be assembled either by overloaded algebra or by the explicit constructor \texttt{MPOLY(n,coef,pow)}. Polynomial matrices are formed by arranging scalar objects into MATLAB arrays or by using helpers such as \texttt{zeros}, \texttt{ones}, \texttt{identity}, and \texttt{diag}. In practice this gives the toolbox two complementary construction paths: a variable-driven path for interactive modeling and a coefficient-exponent path for importing algebraic data from another routine. Even in the direct-constructor path, the resulting object still carries variable metadata, so users can inspect \texttt{p.vars.name} and \texttt{p.vars.id}; by default these names are initialized as \texttt{x1}, \texttt{x2}, \ldots, \texttt{xn}.

For example, consider the scalar polynomial
\[
p(x_1,x_2,x_3) = x_1^3 + 2x_1x_2 - x_3 + 1.
\]
The following transcript illustrates a compact scalar workflow:
\begin{verbatim}
>> x = MPOLY.mpolyvars(3,'x');
>> p = x(1)^3 + 2*x(1)*x(2) - x(3) + 1;
>> p.degree
>> p.sdisp('%.2f')
\end{verbatim}
The expected semantics are:
\begin{verbatim}
p.degree = 3
p = 1.00 -1.00*x3 +2.00*x1*x2 +1.00*x1^3
\end{verbatim}
In addition to display functions such as \texttt{disp} and \texttt{sdisp}, Polylab supports structural inspection through \texttt{degree}, \texttt{coefficients}, \texttt{mono}, and \texttt{iszero}. Here \texttt{disp} reports object size and monomial counts, whereas \texttt{sdisp(format)} prints the actual polynomial expression with user-selected coefficient formatting. Basis-generation and structured-construction helpers such as \texttt{monolist}, \texttt{quadprod}, and \texttt{transmatconvex} are especially useful in optimization-oriented workflows where users assemble polynomial bases and quadratic forms repeatedly. Constant-array builders such as \texttt{zeros}, \texttt{ones}, and \texttt{identity} provide the polynomial analogue of MATLAB's standard matrix constructors, \texttt{iszero} tests structural zero rather than numerical near-zero, and \texttt{diag} and \texttt{trace} follow the usual matrix semantics for diagonal extraction and diagonal summation.

As a matrix-valued example, let
\[
T(x) =
\begin{bmatrix}
x_1+1 & x_2 \\
x_3   & x_1+x_2
\end{bmatrix}.
\]
A representative helper-level session is
\begin{verbatim}
>> I = MPOLY.identity(3,2);
>> T = [x(1)+1, x(2); x(3), x(1)+x(2)];
>> trace(T)
>> numel(MPOLY.monolist(3,2))
\end{verbatim}
with the interpretation
\begin{verbatim}
trace(T) = 1.000000e+00 +2.000000e+00*x1 +1.000000e+00*x2
numel(MPOLY.monolist(3,2)) = 10
\end{verbatim}
More explicitly, \texttt{monolist(n,d)} returns the column vector of all monomials
\[
\{x^\alpha : \alpha \in \mathbb{N}^n,\ |\alpha| \le d\},
\]
arranged in Polylab's internal monomial ordering. Thus, when \(n=3\) and \(d=2\), the basis is
\[
[1,\ x_1,\ x_2,\ x_3,\ x_1^2,\ x_1x_2,\ x_2^2,\ x_1x_3,\ x_2x_3,\ x_3^2]^\top,
\]
which explains both the output length and the combinatorial count \(\binom{n+d}{d}=\binom{5}{2}=10\). This basis generator is one of the package's main building blocks for approximation, moment, and sum-of-squares style workflows.

\subsection{Variable Identity and Safe Mixed-Variable Algebra}
The most important semantic change in Version~3.0 is the move from purely positional variables to explicit variable metadata. In Polylab, each polynomial object carries a variable record with two pieces of information: a stable internal identity \texttt{vars.id} and a human-readable name \texttt{vars.name}. When two polynomial objects are combined, the toolbox aligns their variable spaces before performing the operation.

To make this concrete, consider
\[
p(x_1,x_2)=x_1^2+2x_2+1,
\qquad
q(x_1,y_1,y_2)=y_1^2-x_1y_2+3.
\]
This design matters whenever expressions are built independently. The following session is representative:
\begin{verbatim}
>> x = MPOLY.mpolyvars(2,'x');
>> y = MPOLY.mpolyvars(2,'y');
>> p = x(1)^2 + 2*x(2) + 1;
>> q = y(1)^2 - x(1)*y(2) + 3;
>> r = simplify(p + q);
\end{verbatim}
The resulting object has four aligned variables. The bundled paper example produces the following output:
\begin{verbatim}
r.vars.name = {x1, x2, y1, y2}
r = 4.000000e+00 +1.000000e+00*y1^2
    +2.000000e+00*x2-1.000000e+00*x1*y2
    +1.000000e+00*x1^2
\end{verbatim}
Without explicit variable alignment, this kind of mixed construction is exactly where many positional toolchains become fragile.

\subsection{Differentiation, Evaluation, and Substitution}
Differential operators are exposed with the same overloaded style as the algebraic operators. For a scalar polynomial, \texttt{diff(p,i)} differentiates with respect to variable \texttt{i}, \texttt{jacobian(p)} returns a gradient row, and repeated Jacobians provide Hessian-like polynomial arrays. Evaluation and substitution are handled by \texttt{eval} and \texttt{subs}.

For the polynomial
\[
p(x_1,x_2,x_3)=x_1^3+2x_1x_2-x_3+1,
\]
the first derivative with respect to \(x_1\) is \(3x_1^2+2x_2\), the gradient is
\[
\nabla p(x)=\bigl[3x_1^2+2x_2,\ 2x_1,\ -1\bigr],
\]
and \(p(1,2,3)=3\).
A representative session is
\begin{verbatim}
>> g1 = diff(p,1);
>> J  = jacobian(p);
>> v  = p.eval([1;2;3]);
\end{verbatim}
with the corresponding interpretation
\begin{verbatim}
g1 = 2*x2 + 3*x1^2
J  = [2*x2 + 3*x1^2, 2*x1, -1]
v  = 3
\end{verbatim}
On the CPU path, double-precision evaluation uses a compiled \texttt{mexeval} kernel by default when it is available. The default mode is \texttt{auto}: it tries the MEX path first and otherwise falls back to the pure MATLAB implementation. This compiled helper is built by the repository installation path, where \texttt{install.m} calls \texttt{compilemex.m}. Users can inspect or override the behavior explicitly:
\begin{verbatim}
>> MPOLY.getEvalBackend()
>> MPOLY.setEvalBackend('matlab');
>> v1 = p.eval([1;2;3]);
>> MPOLY.setEvalBackend('auto');
>> v2 = p.eval([1;2;3]);
>> MPOLY.setEvalBackend('mex');
>> v3 = p.eval([1;2;3]);
\end{verbatim}
In the bundled regression test, the MATLAB, AUTO, and forced-MEX modes agree to machine precision on this example. This interface is useful when portability matters more than raw speed, or when users want to benchmark the compiled and interpreted CPU paths separately.

For matrix-valued polynomial expressions, the same operators act entrywise. This is important for Jacobian and Hessian assembly in the affine-normal module, but it is equally useful in more classical symbolic-numeric workflows such as polynomial approximation, sensitivity analysis, and optimization preprocessing. More generally, the overloaded arithmetic follows MATLAB's established distinction between elementwise and matrix algebra: \texttt{plus}, \texttt{minus}, \texttt{times}, \texttt{rdivide}, and \texttt{power} act entrywise, whereas \texttt{mtimes}, \texttt{mrdivide}, and \texttt{mpower} follow matrix semantics. The reduction operators \texttt{sum} and \texttt{prod} act along the selected dimension of a polynomial array.

For the polynomial matrix
\[
A(x)=
\begin{bmatrix}
x_1 & 1 \\
0   & x_2
\end{bmatrix},
\]
the elementwise square is
\[
A(x)^{\circ 2}=
\begin{bmatrix}
x_1^2 & 1 \\
0 & x_2^2
\end{bmatrix},
\]
whereas the matrix square is
\[
A(x)^2=
\begin{bmatrix}
x_1^2 & x_1+x_2 \\
0 & x_2^2
\end{bmatrix}.
\]
Likewise, the calls \texttt{sum(A,1)} and \texttt{prod(A,2)} return the column sums and row products, namely
\[
\texttt{sum(A,1)} = [x_1,\ 1+x_2],
\qquad
\texttt{prod(A,2)} = [x_1;\ 0].
\]
A short session makes these differences explicit:
\begin{verbatim}
>> A = [x(1), 1; 0, x(2)];
>> A_elem = A.^2;
>> A_mat  = A^2;
>> sA = sum(A,1);
>> pA = prod(A,2);
\end{verbatim}
with the corresponding meanings
\begin{verbatim}
A.^2 = [x1^2, 1; 0, x2^2]
A^2  = [x1^2, x1+x2; 0, x2^2]
sum(A,1)  = [x1, 1+x2]
prod(A,2) = [x1; 0]
\end{verbatim}
The same MATLAB-style distinction applies to division: \texttt{mrdivide} is currently intended for division by numeric scalars, while \texttt{rdivide} performs elementwise division by same-size numeric arrays.

\subsection{Coefficient Access, Substitution, and Optimization-Oriented Helpers}
Several functions become especially useful once a polynomial has been constructed and differentiated. For coefficient inspection, for example,
\[
u(x_1,x_2)=3x_1^2+5x_2
\]
\begin{verbatim}
>> u = MPOLY(2,[3;5],[2 0;0 1]);
>> u.vars.name
>> [c,m] = u.coefficients();
>> c
>> m.sdisp('%.0f')
>> u.mono(1).sdisp('%.0f')
\end{verbatim}
returns
\begin{verbatim}
u.vars.name = {x1, x2}
c = [3; 5]
m = [x1^2; x2]
u.mono(1) = x1^2
\end{verbatim}
This also illustrates the direct-constructor behavior: when a polynomial is created from \texttt{(n,coef,pow)} rather than from \texttt{mpolyvars}, Polylab still assigns a consistent default variable record. The corresponding names are available through \texttt{u.vars.name}, and the aligned internal identifiers can likewise be inspected through \texttt{u.vars.id}. For workflows that start from named modeling variables and later combine independently created expressions, the \texttt{mpolyvars} route remains the recommended entry point. The related function \texttt{mono(idx)} therefore exposes basis terms directly, while \texttt{coefficients} keeps coefficient vectors and monomial bases aligned in a form that is convenient for regression, optimization, and moment-style constructions.

If we substitute \(x_1=z_1+1\) and \(x_2=z_2\), then
\[
u(z_1+1,z_2)=3(z_1+1)^2+5z_2
=3+6z_1+3z_1^2+5z_2.
\]
Substitution keeps the same object-oriented interface while replacing variables by new polynomial expressions:
\begin{verbatim}
>> z  = MPOLY.mpolyvars(2,'z');
>> us = u.subs([z(1)+1; z(2)]);
>> us.sdisp('%.0f')
\end{verbatim}
with output
\begin{verbatim}
us = 3 + 6*z1 + 3*z1^2 + 5*z2
\end{verbatim}
so that subsequent simplification, evaluation, export, or backend conversion can proceed on the substituted expression without leaving the Polylab data model.

For structured quadratic construction, Polylab can build
\[
q_f(x)=b(x)^\top Q\,b(x),
\qquad
Q=\begin{bmatrix}2&1\\1&3\end{bmatrix},
\qquad
b(x)=\begin{bmatrix}x_1\\x_2x_3\end{bmatrix},
\]
directly from \(Q\) and \(b\), without manual expansion.
Optimization-oriented helpers support structured constructions that would otherwise be tedious to assemble manually. With
\begin{verbatim}
>> Q  = [2 1; 1 3];
>> b  = [x(1); x(2)*x(3)];
>> qf = MPOLY.quadprod(Q,b);
\end{verbatim}
Polylab forms the quadratic expression
\begin{verbatim}
qf = 2*x1^2 + 2*x1*x2*x3 + 3*x2^2*x3^2
\end{verbatim}
without hand-expanding all pairwise products. Likewise,
\begin{verbatim}
>> [P,Ph,B,mncoefs] = MPOLY.transmatconvex(3,2);
\end{verbatim}
returns basis-transformation data for homogeneous degree-two polynomial models in three variables. The output \texttt{B} enumerates all ways of distributing the total degree among the variables, so each row of \texttt{B} corresponds to one degree pattern such as \((2,0,0)\), \((1,1,0)\), or \((0,0,2)\). The vector \texttt{mncoefs} stores the associated multinomial coefficients, \texttt{Ph} is the raw transformation matrix built from these degree patterns, and \texttt{P} is the coefficient-weighted version actually used in convex homogeneous polynomial representations. In this example \texttt{P} and \texttt{Ph} are \(6 \times 6\) matrices because there are six monomials of total degree two in three variables. This function is therefore best understood as an infrastructure routine for changing between monomial-coefficient coordinates and a structured convex basis, rather than as a pointwise polynomial evaluation command.

\subsection{Backend Conversion and Precision Control}
A second major software concern is backend portability. In Polylab, a CPU polynomial can be sent to the legacy GPU or high-performance GPU-oriented backend without rewriting the model:
\begin{verbatim}
>> p_gpu = MPOLY_GPU.gpulize(p);
>> p_hp  = MPOLY_HP.gpulize(p);
>> p_cpu = MPOLY_HP.cpulize(p_hp);
\end{verbatim}
This is intended for real workflows, not just demos. The tutorial script uses exactly this pattern to verify consistency across backends.

The CPU and HP classes also participate in global precision control:
\begin{verbatim}
>> polylab_precision('set','single');
>> MPOLY.getPrecision()
>> MPOLY_HP.getPrecision()
>> xt = MPOLY.mpolyvars(2,'x');
>> class((xt(1)^2 + 1).coef)
\end{verbatim}
with the expected response
\begin{verbatim}
MPOLY.getPrecision()    = single
MPOLY_HP.getPrecision() = single
class(...)              = single
\end{verbatim}
Precision changes affect subsequent object creation and therefore let a user explore precision-speed tradeoffs without rewriting expressions or constructors. In Polylab Version~3.1, the precision-control interface is implemented for the CPU and HP classes, which is also where the new affine-normal functionality is exposed.

\subsection{LaTeX Export and Reconstruction}
Polylab supports LaTeX export on all main classes and CPU-side reconstruction through \texttt{MPOLY.fromlatex}. This gives the toolbox a useful round-trip path for reports, paper figures, regression tests, and CPU-side parsing of externally edited formulas. For example,
\[
A(x)=
\begin{bmatrix}
x_1^2+x_2 & 2x_1x_2 \\
x_1+1     & x_1-x_2
\end{bmatrix}.
\]
The intended workflow is simple:
\begin{verbatim}
>> A = [x(1)^2 + x(2), 2*x(1)*x(2);
        x(1) + 1,      x(1) - x(2)];
>> s  = latex(A,'x','%.6e');
>> A2 = MPOLY.fromlatex(s);
\end{verbatim}
In our paper example, the re-exported LaTeX string differs from the original only by benign term ordering, and the coefficient-exponent structure is preserved. This is useful for reports, notebooks, regression tests, and CPU reconstruction followed by GPU/HP conversion.

\section{Interoperability with Other MATLAB Ecosystems}
Polylab already contains explicit interfaces to YALMIP and SOSTOOLS, and it also admits a lightweight bridge to the Symbolic Math Toolbox through LaTeX strings.

\subsection{YALMIP}
YALMIP is a widely used modeling framework in MATLAB~\cite{lofberg2004yalmip}. Polylab provides both export and import for YALMIP polynomial expressions, which lets users keep Polylab as the polynomial-construction layer while handing optimization modeling to YALMIP. For the polynomial vector
\[
P(x)=
\begin{bmatrix}
x_1+x_2x_3-2x_3^2 \\
x_1+x_2+x_3
\end{bmatrix},
\]
a representative example is:
\begin{verbatim}
>> xv = sdpvar(3,1);
>> py = mpoly2yalmip(P,xv);
>> Pr = yalmip2mpoly(py,xv);
\end{verbatim}
The paper example records the displayed YALMIP expressions as
\begin{verbatim}
sdisplay(py){1} = xv(1)-2*xv(3)^2+xv(2)*xv(3)
sdisplay(py){2} = xv(1)+xv(2)+xv(3)
round-trip structure check = 1
\end{verbatim}
This round-trip is available for the CPU class, and GPU/HP conversion can be applied afterward. Dedicated import functions are also provided for the GPU and HP backends.

\subsection{SOSTOOLS}
SOSTOOLS is a standard toolbox for sum-of-squares programming in MATLAB~\cite{papachristodoulou2013sostools}. Polylab currently provides export to SOSTOOLS objects for the same polynomial vector \(P(x)\):
\begin{verbatim}
>> xs = mpvar('x',[3,1]);
>> ps = mpoly2sostools(P,xs);
\end{verbatim}
In our example, the exported object has class \texttt{polynomial} and size \([2,1]\). At present the interface is one-way: Polylab exports to SOSTOOLS, but the reverse conversion is not implemented.

\subsection{Symbolic Math Toolbox}
There is no dedicated direct adapter between Polylab and MATLAB's Symbolic Math Toolbox in Polylab Version~3.1. However, LaTeX export provides a practical bridge for some workflows. For example,
\[
p(x_1,x_2,x_3)=x_1+2x_2+x_3+1
\]
may be exported and then converted into a Symbolic Math Toolbox expression through a lightweight string rewrite:
\begin{verbatim}
>> s = latex(p,'x','%.6e');
>> s_sym = ... % convert LaTeX-style x_{i} to x_i syntax
>> str2sym(s_sym)
\end{verbatim}
This example shows that a simple scalar polynomial can already be converted to a symbolic expression through a lightweight string transformation. The current route is a practical bridge rather than a full native symbolic adapter.

\section{Affine-Normal Computation in Polylab}
Version~3.1 extends \texttt{MPOLY} and \texttt{MPOLY\_HP} with affine-normal computation. From the optimization viewpoint, affine-normal directions underlie Yau's Affine Normal Descent (YAND), where they serve as affine-invariant geometry-aware search directions with strong robustness to anisotropic scaling and Newton-like local behavior~\cite{niu2026yand}. The implementation follows the computational framework developed in \emph{Affine Normal Directions via Log-Determinant Geometry: Scalable Computation under Sparse Polynomial Structure}~\cite{niu2026affinenormal}. For a regular level set \(\Sigma_c = \{x \in \mathbb{R}^n : f(x)=c\}\) with \(\nabla f(x) \neq 0\), the affine normal is the distinguished equiaffine-invariant transverse direction. In contrast to the Euclidean normal, which is determined by first derivatives alone, the affine normal encodes second- and third-order local geometry of the level set. This makes it relevant both in affine differential geometry and in geometry-aware direction construction for polynomial optimization, continuation, and descent methods.

At a point \(x\), let \(g = \nabla f(x)\) and \(e = g / \lVert g \rVert\). Choose an orthonormal frame \(Q = [Q_T, e]\) whose last axis aligns with the gradient. In the log-determinant formulation one works in the locally convex regime where the tangent Hessian block is nonsingular, typically positive definite after orientation; the implementation further stabilizes the tangent solves by a small user-controlled regularization. The tangent Hessian block and mixed term are then
\[
H_T = Q_T^\top \nabla^2 f(x) Q_T, \qquad h = Q_T^\top \nabla^2 f(x) e.
\]
The correction term that distinguishes the affine normal from a purely second-order direction can be written as the tangent gradient of \(\log \det H_T\), namely
\[
a_i = \mathrm{tr}\!\left(H_T^{-1} \partial_{t_i} H_T\right), \qquad i=1,\dots,n-1,
\]
where \(\partial_{t_i} H_T\) is represented through third directional derivatives of \(f\). Polylab returns the transverse affine-normal direction \(\nu_{\mathrm{aff}}\) in the form
\[
\nu_{\mathrm{aff}} = Q\begin{bmatrix}\tau \\ -1\end{bmatrix}, \qquad
\tau = H_T^{-1}\!\left(h - \frac{\lVert g \rVert}{n+1} a\right).
\]
The \texttt{ad} route evaluates this expression directly from repeated differentiation in the polynomial model. MF-logDet-Exact computes the same correction without explicitly materializing the full third-order tensor, whereas MF-logDet-Stochastic replaces exact trace evaluation by Hutchinson-style estimation together with Krylov linear solves~\cite{hutchinson1989trace}. Accordingly, the AD route serves as a transparent reference path, while the matrix-free routes are the production methods for larger sparse problems. In the toolbox, the three user-facing methods are:
\begin{itemize}
\item \texttt{ad}: an automatic-differentiation reference route;
\item \texttt{MF-logDet-Exact}: the exact matrix-free log-determinant method;
\item \texttt{MF-logDet-Stochastic}: the stochastic matrix-free log-determinant method.
\end{itemize}

At the MATLAB level, the option strings are \texttt{'ad'}, \texttt{'mf-exact'}, and \texttt{'mf-stochastic'}. In Polylab Version~3.1, omitting \texttt{opts.method} selects \texttt{'mf-exact'} as the package default; the paper benchmarks nevertheless pass all method-specific options explicitly. The paper uses the names MF-logDet-Exact and MF-logDet-Stochastic to stay consistent with the terminology of the companion algorithm paper.

\begin{table}
\caption{User-facing affine-normal options, defaults, and outputs.}
\label{tab:affineopts}
\begin{tabular}{p{0.26\linewidth}p{0.20\linewidth}p{0.43\linewidth}}
\toprule
Field or output & Default or paper value & Meaning \\
\midrule
\texttt{opts.method} & \texttt{'mf-exact'} default; \texttt{'ad'} or \texttt{'mf-stochastic'} optional & selects the AD baseline, exact log-determinant route, or stochastic log-determinant route \\
\texttt{opts.regularization} & \texttt{1e-6} default; \(10^{-8}\) in the paper's accuracy/exact tests & regularization used in linear solves inside the log-determinant computation \\
\texttt{opts.num\_probes} & \texttt{8} default; \texttt{64} in the paper's accuracy tables & Hutchinson probe count for MF-logDet-Stochastic \\
\texttt{opts.krylov\_tol}, \texttt{opts.krylov\_maxit} & \texttt{1e-4}, \texttt{40} defaults; \texttt{80} iterations in the paper's accuracy tables & Krylov stopping parameters for MF-logDet-Stochastic \\
\texttt{opts.backend\_mode} & \texttt{'auto'} on \texttt{MPOLY\_HP} & lets MPOLY-HP choose the appropriate execution path \\
\texttt{opts.assume\_nonzero\_x} & logical flag & skips conservative checks when the evaluation point is known to be safe \\
\texttt{d} & numeric or \texttt{gpuArray} vector & affine-normal direction at the evaluation point \\
\texttt{info} & structure & diagnostics such as method name, gradient norm, and work counters \\
\bottomrule
\end{tabular}
\Description{A table summarizing the user-facing option structure, package defaults, and outputs for affine-normal computation in Polylab.}
\end{table}

For the quartic example
\[
f(x)=x_1^4+2x_1^2x_2^2+x_2^4+x_3^2+3x_4+1,
\qquad
x_0=(0.3,0.2,0.4,0.1)^\top,
\]
Polylab therefore exposes affine-normal computation as a direct class method:
\begin{verbatim}
>> x = MPOLY.mpolyvars(4,'x');
>> p = x(1)^4 + 2*x(1)^2*x(2)^2 + x(2)^4 + x(3)^2 + 3*x(4) + 1;
>> x0 = [0.3; 0.2; 0.4; 0.1];
>> [d,info] = MPOLY.affineNormalDirection(...
       p,x0,struct('method','mf-exact','regularization',1e-8));
\end{verbatim}
The generated example output is:
\begin{verbatim}
d = [-3.681054e+00; -2.454036e+00;
     -1.013790e-09; -7.603422e-01]
info.method = mf-exact
info.grad_norm = 3.110491e+00
info.hv_evals = 4
info.third_evals = 3
\end{verbatim}
The same polynomial can be sent to the HP backend and evaluated through
\begin{verbatim}
>> pg = MPOLY_HP.gpulize(p);
>> [dg,infog] = MPOLY_HP.affineNormalDirection(...
       pg,gpuArray(x0),struct('method','mf-exact', ...
       'regularization',1e-8,'backend_mode','auto', ...
       'assume_nonzero_x',true));
\end{verbatim}
In our example, the CPU and HP directions agree to the displayed precision.

\section{Experimental Evaluation}
\subsection{Setup}
All results reported here were generated from repository scripts. Functional coverage comes from the repository backend tests, tutorial script, affine-normal test, and dedicated precision benchmarks. For the paper itself we prepared a compact driver, \path{paper/scripts/run_paper_experiments.m}, together with companion scripts \path{paper/scripts/run_paper_additional_examples.m}, \path{paper/scripts/run_backend_regime_benchmarks.m}, and \path{paper/scripts/run_stochastic_stability_benchmark.m}. The scripts fix the random seed to \texttt{20260321}, \texttt{20260322}, or the per-seed values recorded in the generated CSV files, write their derived tables and transcript files into \path{paper/data}, and export the publication figures into \path{paper/figures}. Small symbolic workloads are averaged over a few repetitions to reduce launch-noise, while the heavier affine-normal and regime-separation routes use one to three repetitions as recorded directly in the scripts and generated CSV files. The experiments were run under MATLAB R2023b on Windows 10 Pro (build 26200) with a 12th Gen Intel(R) Core(TM) i9-12900K CPU (24 logical processors) and an NVIDIA GeForce RTX~4090. For affine-normal tests, the HP backend uses \texttt{backend\_mode = auto}. The exact method uses regularization \(10^{-8}\). For the affine-normal accuracy and precision tables, the stochastic method uses an accuracy-oriented setting with 64 probes and Krylov iteration limit 80; only Figure~\ref{fig:stochastic} switches to speed-oriented stochastic parameters. Table~\ref{tab:benchmeta} records the benchmark scales and repetition counts. Code will be made available upon publication.

The affine-normal routines implemented in Polylab follow the MF-logDet framework of Niu, Sheshmani, and Yau~\cite{niu2026affinenormal}. This section reports implementation-level correctness checks, backend-dependent runtime behavior, and representative usage regimes of the current Polylab interface. Broader optimization motivation and convergence theory for affine-normal descent are developed in the YAND paper~\cite{niu2026yand}, while broader methodological validation, complexity analysis, and extended polynomial experiments for the MF-logDet computational framework are reported in the companion paper~\cite{niu2026affinenormal}. Since those experiments were not conducted through the present Polylab interface, we cite them as evidence for the underlying algorithmic framework rather than as direct software benchmarks.

\begin{table}
\caption{Benchmark workload summary.}
\label{tab:benchmeta}
\small
\begin{tabular}{p{0.17\linewidth}p{0.23\linewidth}p{0.48\linewidth}}
\toprule
Item & Family & Scale \\
\midrule
Fig.~\ref{fig:corebench}, Table~\ref{tab:coreops} & Lightweight core ops & Five-variable scalar objects plus 2\(\times\)2 and 3\(\times\)3 polynomial matrices; \texttt{monolist(6,4)}; simplify uses \((\sum \mathcal{M}_{5,3})^3\) with 56 base monomials; 2--20 repetitions. \\
Fig.~\ref{fig:simplifyregime} & Heavy simplify & \(p_d(x)=\sum_{|\alpha|\le d}x^\alpha\) in five variables, with \(d\in\{2,3,4\}\) and 21, 56, 126 base monomials; benchmark is \(\operatorname{simplify}(p_d^3)\); 1--2 repetitions. \\
Table~\ref{tab:accuracy} & Affine accuracy & One sparse quartic in six variables with 17 monomials at one sample point; AD reference; stochastic uses the accuracy-oriented setting \((q=64,\ \texttt{krylov\_maxit}=80)\). \\
Fig.~\ref{fig:exact} & Exact scaling & Sparse quartics with support size 3 and \(m=4d\), for \(d\in\{20,40,80,120,180,260\}\); single and double precision; 3 repetitions. \\
Fig.~\ref{fig:stochastic} & Stochastic crossover & Sparse quartics with support size 2 and \(m=3d\), for \(d\in\{80,120,180,260,360,500\}\); exact repetitions 1, stochastic repetitions 3; speed-oriented setting \((q=1,\ \texttt{regularization}=5\times 10^{-4},\ \texttt{krylov\_tol}=2\times 10^{-2},\ \texttt{krylov\_maxit}=6)\). \\
Table~\ref{tab:precision} & Precision tests & Core: random degree-4 polynomials with \(n=8\), 300 monomials, and 3\(\times\)3 matrix replication. Affine: one eight-variable sparse quartic with 23 monomials. 2 repetitions. \\
\bottomrule
\end{tabular}
\Description{A table summarizing the workload families, scales, and repetition counts for the benchmark figures and tables in the paper.}
\end{table}

\subsection{Lightweight Core Symbolic Benchmarks}
\begin{figure}
\includegraphics[width=\linewidth]{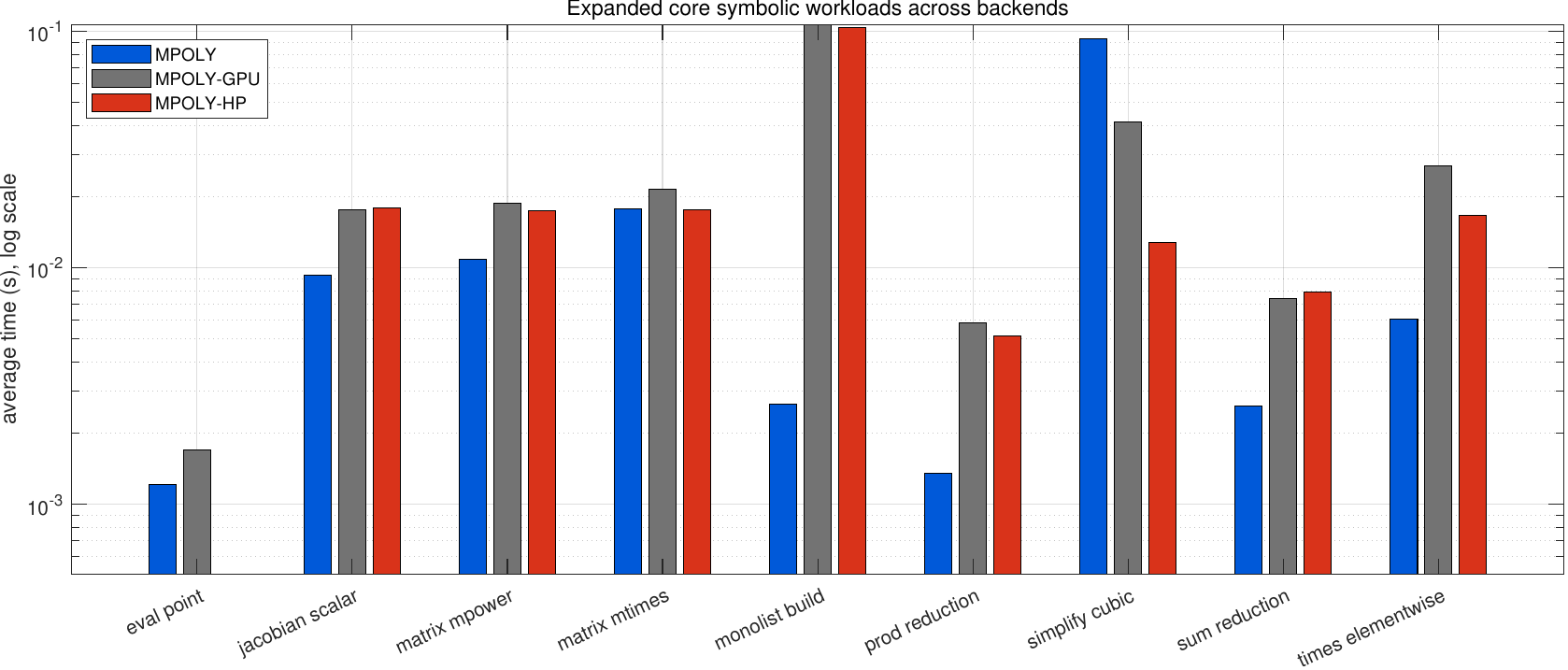}
\caption{Expanded lightweight core symbolic workloads across MPOLY, MPOLY-GPU, and MPOLY-HP. The workloads use 5-variable scalar or small polynomial-matrix objects; the vertical axis is logarithmic.}
\Description{A grouped bar chart on a logarithmic scale comparing lightweight core polynomial operations such as monolist construction, symbolic simplify, elementwise multiplication, matrix multiplication, matrix power, reduction operators, jacobian construction, and pointwise evaluation across MPOLY, MPOLY-GPU, and MPOLY-HP.}
\label{fig:corebench}
\end{figure}

Figure~\ref{fig:corebench} expands the core benchmark beyond \texttt{mtimes} and \texttt{mpower} and serves as a lightweight-to-medium interactive workload study. The suite covers basis construction (\texttt{monolist}), symbolic expansion and simplification, elementwise multiplication, matrix multiplication, matrix powers, reduction operators (\texttt{sum} and \texttt{prod}), differentiation through \texttt{jacobian}, and pointwise evaluation through \texttt{eval}. Problem sizes are summarized in Table~\ref{tab:benchmeta}. The vertical axis is logarithmic, and Table~\ref{tab:coreops} complements the figure with exact timings.

\begin{table}
\caption{Expanded backend comparison for representative core operations (seconds).}
\label{tab:coreops}
\small
\begin{tabular}{lrrrr}
\toprule
Task & MPOLY & MPOLY-GPU & MPOLY-HP & Fastest \\
\midrule
\texttt{monolist} build & 0.0023 & 0.1014 & 0.0972 & MPOLY \\
\texttt{simplify} after cubic expansion & 0.0954 & 0.0431 & 0.0134 & MPOLY-HP \\
Elementwise \texttt{times} & 0.0050 & 0.0261 & 0.0172 & MPOLY \\
Matrix \texttt{mtimes} & 0.0149 & 0.0207 & 0.0209 & MPOLY \\
Matrix \texttt{mpower} & 0.0051 & 0.0187 & 0.0149 & MPOLY \\
Reduction \texttt{sum} & 0.0028 & 0.0077 & 0.0081 & MPOLY \\
Reduction \texttt{prod} & 0.0013 & 0.0038 & 0.0051 & MPOLY \\
\texttt{jacobian} & 0.0056 & 0.0172 & 0.0157 & MPOLY \\
\texttt{eval} at one point & 0.0010 & 0.0022 & 0.0006 & MPOLY-HP \\
\bottomrule\end{tabular}
\Description{A table comparing runtime in seconds for representative core polynomial operations across MPOLY, MPOLY-GPU, and MPOLY-HP, together with the fastest backend for each task.}
\end{table}

This broader view gives a more nuanced backend picture. On lightweight-to-medium tasks, the CPU implementation remains the right default: it is fastest on basis generation, elementwise arithmetic, matrix multiplication and powers, reductions, and Jacobian construction. MPOLY-HP still delivers the best pointwise \texttt{eval} time in this suite, roughly \(1.63\times\) faster than MPOLY on the tested one-point evaluation workload, and remains the clearest winner for the included cubic expansion-and-simplify workload, where it is about \(7.1\times\) faster than MPOLY and about \(3.2\times\) faster than MPOLY-GPU. Figure~\ref{fig:corebench} is not a universal backend ranking, however. It measures low-overhead primitives, whereas the GPU-oriented backends were designed primarily for workloads dominated by large intermediate expansions, repeated-term consolidation, and affine-geometric linear algebra.

Not every function in Table~\ref{tab:featurecatalog} is benchmarked separately. Display and formatting functions such as \texttt{disp}, \texttt{sdisp}, and \texttt{latex} are dominated by I/O and string-generation effects; simple structural queries such as \texttt{degree}, \texttt{coefficients}, \texttt{mono}, \texttt{trace}, and \texttt{iszero} are too lightweight to be informative in backend timing plots; and CPU-only or backend-asymmetric paths such as \texttt{fromlatex} and the legacy GPU substitution path are better documented through usage examples than through raw timing comparisons. Figure~\ref{fig:corebench} and Table~\ref{tab:coreops} therefore focus on the computationally meaningful operations that are both central to modeling workflows and comparable across the three backends.

\subsection{Reduction-Heavy Simplification Regime}
\begin{figure}
\includegraphics[width=\linewidth]{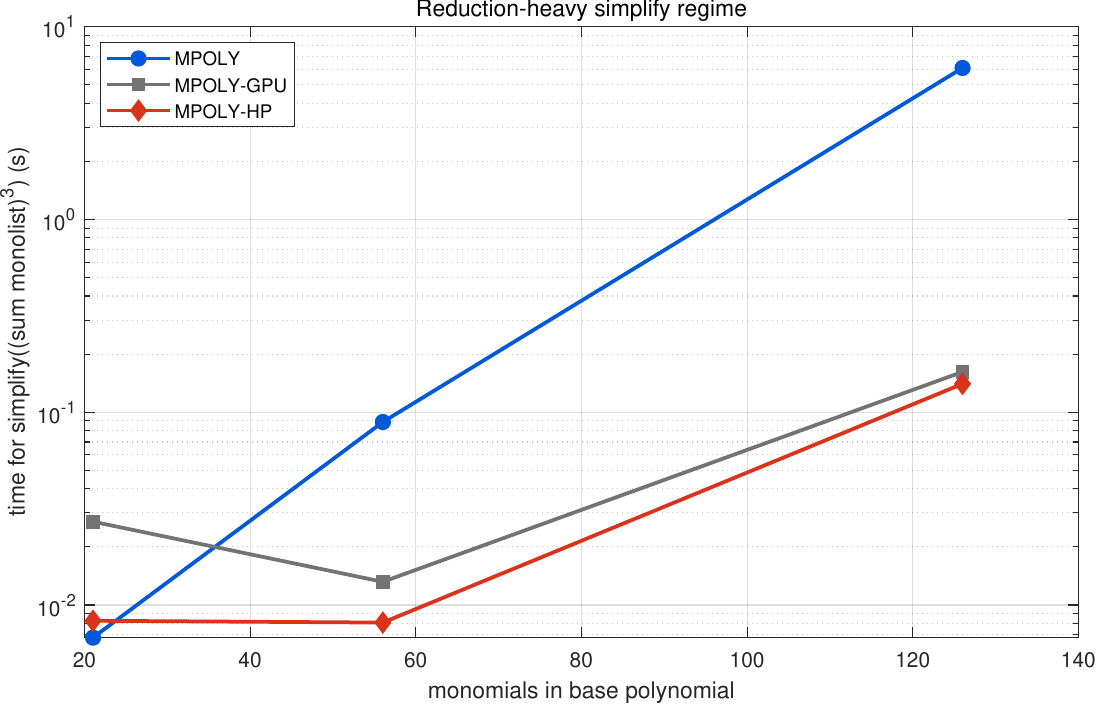}
\caption{Backend crossover on a reduction-heavy simplify workload. The test measures \(\operatorname{simplify}(p_d^3)\) for \(p_d=\sum_{|\alpha|\le d}x^\alpha\) in five variables.}
\Description{A logarithmic line plot of simplify runtime versus monomial count in the base polynomial for MPOLY, MPOLY-GPU, and MPOLY-HP. The CPU curve is competitive at the smallest size but grows much faster, while the GPU-oriented backends become decisively faster as the base polynomial becomes denser.}
\label{fig:simplifyregime}
\end{figure}

Figure~\ref{fig:simplifyregime} isolates a workload class for which the GPU-oriented backends are genuinely valuable: dense symbolic expansion followed by monomial consolidation. The benchmark evaluates
\[
p_d(x)=\sum_{|\alpha|\le d} x^\alpha,\qquad f_d(x)=\operatorname{simplify}(p_d(x)^3)
\]
in five variables, with the horizontal axis reporting the number of monomials in the base polynomial \(p_d\). Here the ranking reverses. At 21 base monomials, MPOLY-HP is already slightly faster than MPOLY, while the legacy GPU path still lags. At 56 monomials, the speedups grow to about \(6.83\times\) for MPOLY-GPU and \(11.44\times\) for MPOLY-HP. At 126 monomials, the CPU path requires about 6.13~s, compared with 0.162~s for MPOLY-GPU and 0.158~s for MPOLY-HP, corresponding to speedups of about \(37.8\times\) and \(38.8\times\), respectively. This is the regime for which the GPU backends were designed: not every primitive operator, but workloads dominated by coefficient aggregation, repeated-term merging, and large intermediate symbolic expansions.

\subsection{Affine-Normal Accuracy}
Table~\ref{tab:accuracy} compares the three affine-normal methods on a representative sparse quartic polynomial in six variables and 17 monomials. The reported case uses the fixed seed recorded in the paper scripts, and the \texttt{ad} route is used as the reference direction.

\begin{table}
\caption{Affine-normal method comparison on a representative quartic polynomial.}
\label{tab:accuracy}
\begin{tabular}{llrrr}
\toprule
Backend & Method & Time (s) & Unit error vs. AD & Angle (deg) \\
\midrule
MPOLY     & AD                    & 0.1093 & 0 & 0 \\
MPOLY     & MF-logDet-Exact       & 0.0068 & $1.99\times 10^{-8}$ & $1.21\times 10^{-6}$ \\
MPOLY     & MF-logDet-Stochastic  & 0.0266 & 0.1088 & 6.24 \\
MPOLY-HP  & AD                    & 0.0738 & 0 & 0 \\
MPOLY-HP  & MF-logDet-Exact       & 0.0021 & $1.99\times 10^{-8}$ & $1.21\times 10^{-6}$ \\
MPOLY-HP  & MF-logDet-Stochastic  & 0.0284 & 0.0290 & 1.66 \\
\bottomrule\end{tabular}
\Description{A table reporting runtime and directional agreement for AD, MF-logDet-Exact, and MF-logDet-Stochastic affine-normal computation on CPU and HP backends. The exact log-determinant method matches the AD reference to nearly machine precision.}
\end{table}

\begin{table}
\caption{Stochastic accuracy stability over 10 probe seeds.}
\label{tab:stochstab}
\small
\begin{tabular}{lrrr}
\toprule
Backend & Mean time (s) & Mean unit error $\pm$ std & Mean angle (deg) $\pm$ std \\
\midrule
MPOLY & 0.0250 & $0.0523 \pm 0.0229$ & $2.99 \pm 1.31$ \\
MPOLY-HP & 0.0238 & $0.0523 \pm 0.0229$ & $2.99 \pm 1.31$ \\
\bottomrule
\end{tabular}
\Description{A table summarizing mean runtime and directional agreement over 10 random probe seeds for the accuracy-oriented stochastic affine-normal method on MPOLY and MPOLY-HP.}
\end{table}

Table~\ref{tab:stochstab} complements the fixed-seed case with a 10-seed summary under the same accuracy-oriented stochastic preset. The mean stochastic angle is about \(2.99^\circ \pm 1.31^\circ\) on both backends, with mean unit error about \(0.0523 \pm 0.0229\); mean runtime stays near \(2.50\times 10^{-2}\)~s on MPOLY and \(2.38\times 10^{-2}\)~s on MPOLY-HP. This indicates that the CPU/HP discrepancy in Table~\ref{tab:accuracy} is largely a probe-realization effect rather than a systematic backend bias. Even so, on this moderate workload \texttt{MF-logDet-Exact} remains both faster and more accurate, so it is still the natural default for high-accuracy affine-normal computation. The speed-oriented stochastic tuning is therefore reserved for the large sparse crossover regime in Figure~\ref{fig:stochastic}.

\subsection{Exact Log-Determinant Scaling}
\begin{figure}
\includegraphics[width=\linewidth]{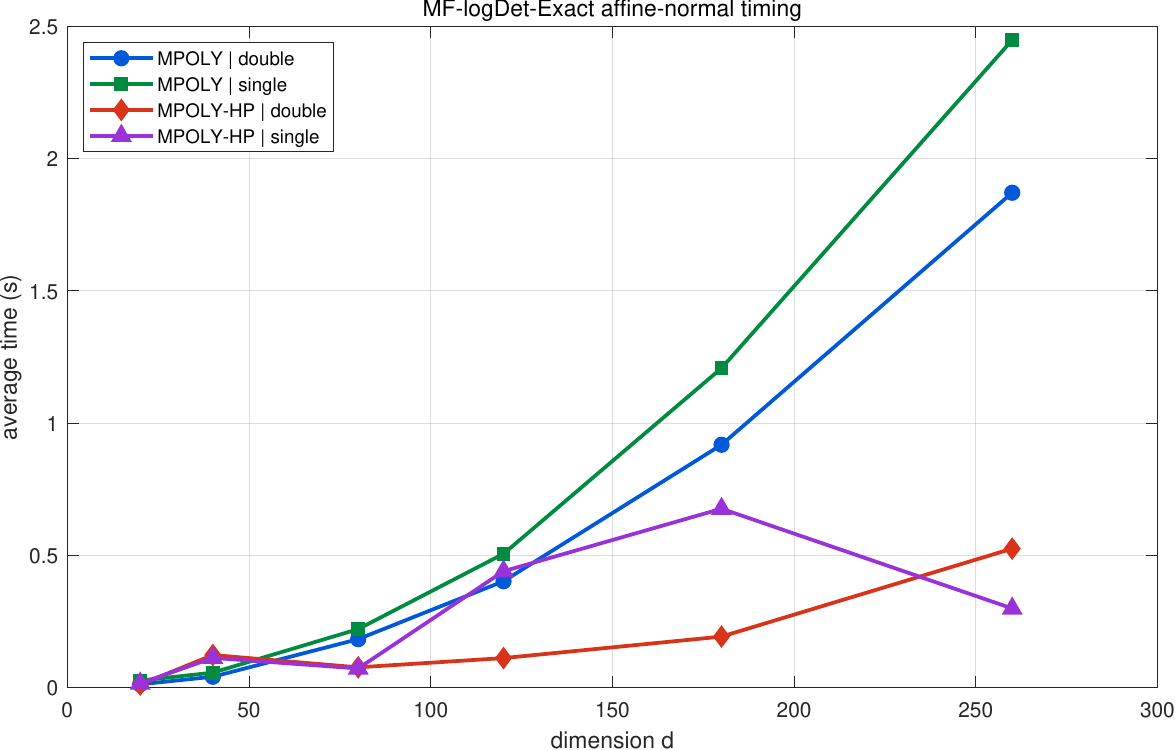}
\caption{MF-logDet-Exact affine-normal timing across dimensions for MPOLY and MPOLY-HP on sparse quartics with support size 3 and \(m=4d\) monomials.}
\Description{A line plot of average affine-normal runtime versus dimension for the exact log-determinant method. MPOLY-HP is faster across most medium and large dimensions in both precisions, although there are small crossover cases at the low-dimensional end.}
\label{fig:exact}
\end{figure}

Figure~\ref{fig:exact} shows how the exact log-determinant method scales across dimension on CPU and HP backends in both single and double precision. The benchmark family uses sparse quartic polynomials with support size 3 and \(m=4d\) monomials, with \(d\in\{20,40,80,120,180,260\}\). Over the tested sweep, MPOLY-HP reduces the average runtime from about 0.571~s to 0.174~s in double precision and from about 0.745~s to 0.269~s in single precision. The curves are not uniformly ordered at the smallest dimensions, where launch overheads and sparse-kernel crossover effects still matter, but MPOLY-HP becomes clearly advantageous through most of the medium-to-large regime.

An interesting implementation detail is that single precision is not uniformly faster on the CPU path. This reflects the current maturity of the double-precision sparse and compatibility paths in MATLAB, the MEX-backed evaluation route, and the toolbox itself.

Table~\ref{tab:precision} gives a more explicit precision comparison on representative workloads and documents the main precision effects of the current implementation.

\begin{table}
\caption{Representative single- versus double-precision timings (seconds).}
\label{tab:precision}
\begin{tabular}{llrrr}
\toprule
Category & Backend/task & Double & Single & Double/Single \\
\midrule
Core & MPOLY eval & 0.0016 & 0.0016 & 0.97 \\
Core & MPOLY simplify & 0.0419 & 0.0051 & 8.26 \\
Core & MPOLY mpower & 8.1507 & 6.1109 & 1.33 \\
Core & MPOLY-HP eval & 0.0230 & 0.0144 & 1.59 \\
Core & MPOLY-HP simplify & 0.0676 & 0.0306 & 2.21 \\
Core & MPOLY-HP mpower & 0.3877 & 0.3047 & 1.27 \\
Affine & MPOLY MF-logDet-Exact & 0.0025 & 0.0076 & 0.33 \\
Affine & MPOLY-HP MF-logDet-Exact & 0.0026 & 0.0020 & 1.31 \\
Affine & MPOLY MF-logDet-Stochastic & 0.0400 & 0.0543 & 0.74 \\
Affine & MPOLY-HP MF-logDet-Stochastic & 0.0379 & 0.0435 & 0.87 \\
\bottomrule\end{tabular}
\Description{A table comparing representative double-precision and single-precision runtimes on core symbolic and affine-normal workloads for MPOLY and MPOLY-HP.}
\end{table}

\subsection{When the Stochastic Method Wins}
\begin{figure}
\includegraphics[width=\linewidth]{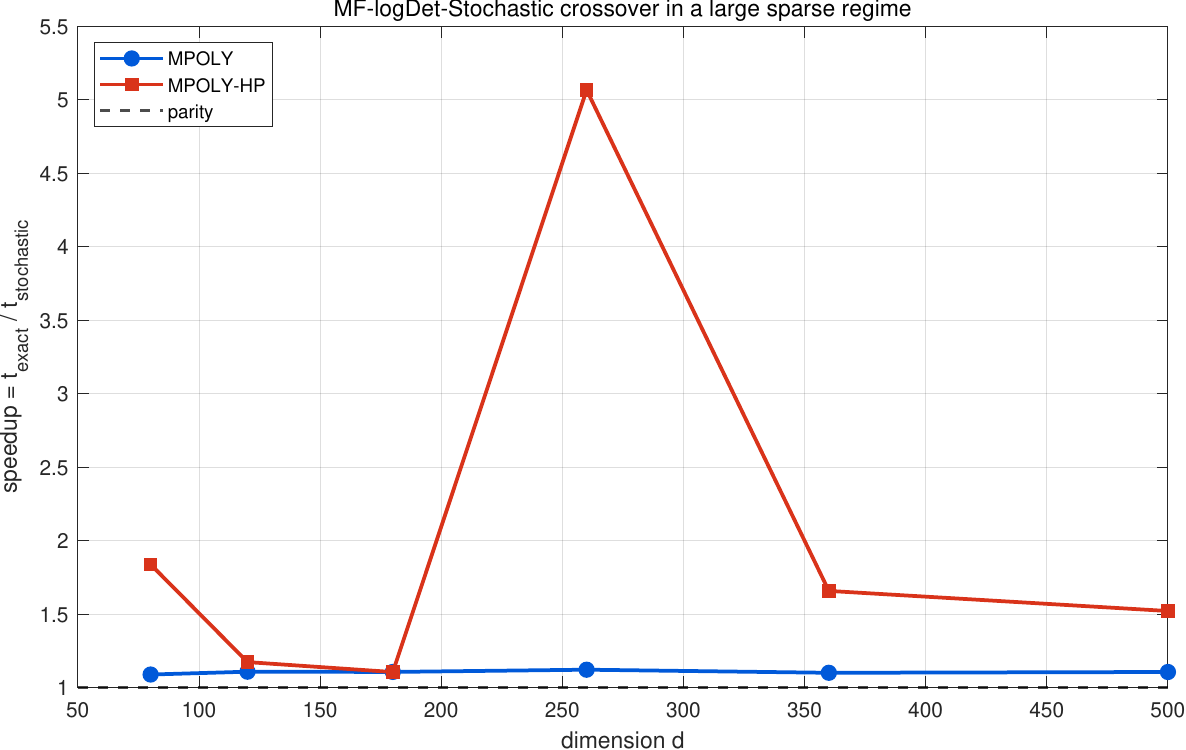}
\caption{Speedup of MF-logDet-Stochastic over MF-logDet-Exact on sparse quartics with support size 2 and \(m=3d\) monomials. Values above one mean that the stochastic method is faster.}
\Description{A line plot of the speedup ratio t_exact divided by t_stochastic versus dimension for MPOLY and MPOLY-HP. Both curves remain above one in the tuned large-scale sparse regime, with the HP backend showing a pronounced mid-range advantage.}
\label{fig:stochastic}
\end{figure}

Figure~\ref{fig:stochastic} reports a benchmark intentionally tuned to reveal the advantage region of the stochastic log-determinant method. The test uses sparse quartics with support size 2 and \(m=3d\) monomials, together with speed-oriented stochastic parameters \((q=1,\ \texttt{regularization}=5\times 10^{-4},\ \texttt{krylov\_tol}=2\times 10^{-2},\ \texttt{krylov\_maxit}=6)\). In this setting, the speedup ratio \(t_{\mathrm{exact}}/t_{\mathrm{stochastic}}\) is above one on both backends over the tested range from 80 to 500 variables.

Table~\ref{tab:stochcases} makes the large sparse crossover concrete on three representative sizes. At 500 variables, the observed speedup is about \(1.11\times\) on MPOLY and \(1.52\times\) on MPOLY-HP. The most pronounced gain in this sweep appears on MPOLY-HP around 260 variables, where the tuned stochastic route is about \(5.07\times\) faster than the exact route. This figure also explains why earlier untuned benchmarks may fail to show any stochastic advantage: the method becomes attractive only when the regime, sparsity pattern, and approximation parameters are aligned with its intended use.

\begin{table}
\caption{Representative large-sparse stochastic crossover cases.}
\label{tab:stochcases}
\small
\begin{tabular}{rrrrrr}
\toprule
\(d\) & \(m\) & MPOLY exact/stoch & MPOLY speedup & MPOLY-HP exact/stoch & MPOLY-HP speedup \\
\midrule
80  & 240  & 0.139 / 0.128 & 1.09 & 0.203 / 0.111 & 1.84 \\
260 & 780  & 1.453 / 1.295 & 1.12 & 1.141 / 0.225 & 5.07 \\
500 & 1500 & 5.285 / 4.776 & 1.11 & 0.788 / 0.518 & 1.52 \\
\bottomrule
\end{tabular}
\Description{A table listing representative dimensions, monomial counts, exact and stochastic runtimes, and speedups for the large sparse stochastic crossover benchmark on MPOLY and MPOLY-HP.}
\end{table}

\section{Discussion and Limitations}
Several conclusions emerge from the software study.

First, Polylab is a layered software system rather than a single algorithm. Its contribution lies in the integration of a polynomial data model, explicit variable semantics, overloaded operations, backend conversion, exchange formats, and geometry-aware computation within one MATLAB interface.

Second, explicit variable identity changes modeling reliability in a substantive way. Once expressions can be created independently, imported from LaTeX or YALMIP, and then recombined, positional variable storage is no longer robust enough on its own. The \texttt{vars.id}/\texttt{vars.name} mechanism addresses this failure mode directly.

Third, performance remains workload dependent, but the present benchmarks make the backend choice much clearer than before. MPOLY is the natural default for interactive modeling and most lightweight primitives, where low overhead matters more than parallel throughput. MPOLY-HP is preferable when the workflow is dominated by large simplification passes, repeated-term consolidation, or affine-normal computation in medium-to-large dimensions. MPOLY-GPU remains useful mainly as a legacy compatibility backend and as a historical baseline, although it can still accelerate some large reduction-heavy workloads.

\begin{table}
\caption{Practical backend-selection guidelines.}
\label{tab:backendguide}
\small
\begin{tabular}{p{0.31\linewidth}p{0.17\linewidth}p{0.39\linewidth}}
\toprule
Scenario & Recommended backend & Reason \\
\midrule
Interactive modeling, basis generation, Jacobians, small matrix algebra, and MEX-backed CPU evaluation & MPOLY & Lowest overhead and best performance on most lightweight primitives in Figure~\ref{fig:corebench} and Table~\ref{tab:coreops}. \\
Large symbolic expansion, \texttt{simplify}-dominated workflows, repeated monomial aggregation, and product-heavy reduction & MPOLY-HP & GPU-native reduction paths; Figure~\ref{fig:simplifyregime} shows the advantage growing from a slight gain at the smallest case to about \(39\times\) at the largest case. \\
Medium- to large-scale affine-normal workloads, exact or stochastic log-det routes & MPOLY-HP & Figures~\ref{fig:exact} and \ref{fig:stochastic} show the clearest backend advantage in geometry-oriented computations. \\
Legacy GPU scripts, historical reproduction, or compatibility with older code paths & MPOLY-GPU & Retained for backward compatibility and still helpful on some heavy simplify workloads, but generally superseded by MPOLY-HP. \\
\bottomrule
\end{tabular}
\Description{A table giving practical guidance for choosing between MPOLY, MPOLY-GPU, and MPOLY-HP based on workload type.}
\end{table}

Fourth, the three affine-normal routes play asymmetrical roles. \texttt{ad} is the transparent reference path, \texttt{MF-logDet-Exact} is the present high-accuracy workhorse, and \texttt{MF-logDet-Stochastic} is a regime-dependent accelerator whose benefits appear only when sparsity, scale, and approximation parameters are aligned.

Fifth, the application record summarized in Table~\ref{tab:applications} suggests that Polylab is already useful outside synthetic benchmarks. Together with the present benchmarks, it shows a package that combines research-prototype flexibility with evidence of practical reuse.

Finally, Polylab Version~3.1 still has clear extension points. A direct Symbolic Math Toolbox adapter, reverse conversion from SOSTOOLS, tensor-derivative interfaces, and more mature precision-specific sparse kernels would all broaden the scope of the toolbox without changing its core design.

\section{Conclusion}
Polylab unifies backend-aware polynomial modeling, variable-safe expression composition, LaTeX round-tripping, ecosystem bridges, precision control, and affine-normal computation in a single MATLAB toolbox. Polylab Version~3.1 combines mature core manipulation routines with geometry-oriented algorithms derived from a recent log-determinant framework. The present benchmarks also make the intended backend roles explicit: MPOLY is the default lightweight engine, whereas MPOLY-HP is the preferred accelerator for reduction-heavy symbolic workloads and affine-geometric computation. Backed both by application use and by reproducible benchmarks, the package serves simultaneously as practical research software and as a platform for further work on scalable symbolic-numeric polynomial computation.

\begin{acks}
Y.-S.\ N.\ was supported by the National Natural Science Foundation of China
(Grant No.\ 42450242) and the Beijing Overseas High-Level Talent Program.
The authors gratefully acknowledge institutional support from the Beijing Institute of Mathematical Sciences and Applications (BIMSA) and Yau Mathematical Sciences Center, Tsinghua University.
\end{acks}

\bibliographystyle{ACM-Reference-Format}
\bibliography{references}

\end{document}